# Microlensing Predictions: Impact of Galactic Disc Dynamical Models

Hongjing Yang (杨弘靖)[1][*] , Shude Mao[1,2] , Weicheng Zang[1] , and Xiangyu Zhang[1]
[1]*Department of Astronomy, Tsinghua University, Beijing 100084, China*
[2]*National Astronomical Observatories, Chinese Academy of Sciences, Beijing 100101, China*

**ABSTRACT**
Galactic model plays an important role in the microlensing field, not only for analyses of individual events but also for statistics of the ensemble of events. However, the Galactic models used in the field varies, and some are unrealistically simplified. Here we tested three Galactic disc dynamic models, the first is a simple standard model that was widely used in this field, whereas the other two consider the radial dependence of the velocity dispersion, and in the last model, the asymmetric drift. We found that for a typical lens mass $M_L = 0.5 M_\odot$, the two new dynamical models predict ∼ 16% or ∼ 5% less long-timescale events (e.g., microlensing timescale $t_E > 300$ days) and ∼ 5% and ∼ 3.5% more short-timescale events ($t_E < 3$ days) than the standard model. Moreover, the microlensing event rate as a function of Einstein radius $\theta_E$ or microlensing parallax $\pi_E$ also shows some model dependence (a few percent). The two new models also have an impact on the total microlensing event rate. This result will also to some degree affect the Bayesian analysis of individual events, but overall, the impact is small. However, we still recommend that modelers should be more careful when choosing the Galactic model, especially in statistical works involving Bayesian analyses of a large number of events. Additionally, we find the asymptotic power-law behaviors in both $\theta_E$ and $\pi_E$ distributions, and we provide a simple model to understand them.

## 1 INTRODUCTION

Since the first detection of Gravitational microlensing event in 1993 (Alcock et al. 1993; Udalski et al. 1993), microlensing has become a powerful tool for probing various objects in the Milky Way, such as extrasolar planets beyond the snow line (Mao & Paczynski 1991; Gould & Loeb 1992), binary stars (e.g., An et al. 2002), isolated black holes (Gould 2000a; Mao et al. 2002; Bennett et al. 2002; Wyrzykowski et al. 2016).

However, the mass measurement of a microlens system is challenging. It requires two observables, i.e., any two quantities out of the angular Einstein radius $\theta_E$, the microlens parallax $\pi_E$ and the apparent brightness of the lens system (for more details, see the Introduction section in Zang et al. 2020). If $\theta_E$ and $\pi_E$ are measured, the mass ($M_L$) of the lens object can be uniquely derived from (Gould 2000b)

$$M_L = \frac{\theta_E}{\kappa \pi_E}, \qquad (1)$$

where $\kappa \equiv 4G/(c^2 \mathrm{AU}) = 8.144\,\mathrm{mas}/M_\odot$ is a constant, AU represents the astronomical unit. Unfortunately, $\theta_E$ and $\pi_E$ are both high-order parameters and usually unobservable, and measurements of the lens light require a ≳ 5 years wait to resolve the lens and source by high-resolution images. This is, however, infeasible for dark lenses such as free-floating planets and black holes.

In most microlensing events, the only available parameter that can be retrieved directly from the observation is what is called the microlensing timescale,

$$t_E = \sqrt{\frac{4GM_L}{c^2}\frac{D_L(D_S - D_L)}{D_S}}\frac{1}{V_t}, \qquad (2)$$

where $V_t$ is the relative tangential velocity of the lens and source. Obviously, the physical parameters $M_L$, $D_L$, $D_S$ and $V_t$ are all degenerate. In this case, the only way to obtain these parameters for an individual event is Bayesian analysis (e.g., Beaulieu et al. 2006) and a Galactic model (GM) as the prior is essential for the analysis.

Since microlensing observations are independent of the lens light, it can detect dark objects like free-floating planets (FFP) and isolated neutron stars (NS) and black holes (BH). For free-floating planet candidates found by Sumi et al. (2011), $t_E$ is the only observable. Some other FFP candidates like OGLE-2016-BLG-1540 (Mróz et al. 2018), OGLE-2012-BLG-1323, and OGLE-2017-BLG-0560 (Mróz et al. 2019), as a result of the strong finite-source effect, their $\theta_E$ are also detectable, while $\pi_E$ is still unknown. These events are believed to be free-floating planets just because of their short timescales ($t_E < 2$ days) and/or small $\theta_E$ ($< 40\mu$as), which lead to high Bayesian posterior probabilities that the lenses are low mass objects. For neutron star and black hole candidates (Wyrzykowski et al. 2016), $\pi_E$ can be obtained from the light curve thanks to their long timescale ($t_E \gtrsim 100$ days) and the Earth motion, but $\theta_E$ remains unknown. The reason that the authors argue they are NS or BH candidates is still their Bayesian posterior. In these identification procedures, a Galactic model is essential to be utilized as prior knowledge.

In addition to individual events, Galactic model also plays a crucial role in microlensing predictions and statistics. For prediction or simulation works, Kiraga & Paczynski (1994) concludes that the microlensing event rate is $\Gamma \approx 12$ per year per $10^6$ stars monitored; Gould (2000a) says about 20% microlensing events are caused by stellar remnants (white dwarfs, neutron stars and black holes); Han (2008) find that if the limiting magnitude is $V_{\rm lim} < 18$, there will be $\Gamma = 23$ near-by (all-sky, without the Galactic bulge) microlensing events per year. For statistical works, Suzuki et al. (2018) shows that microlensing results challenge the core accretion planet formation theory. All these works are based or partly based on a Galactic model (whatever it is).

[*] E-mail: yang-hj19@mails.tsinghua.edu.cn

Therefore, the Galactic model is strongly connected with not only analysis of individual events but also prediction and statistics of ensemble events. However, up to date, the models used in this field are diverse, some are simple while some are complex. For example, in Zhu et al. (2017) (Zhu17), the Galactic model only contains two components, a disc and a bulge (bar). The dynamics is also simple, in the bulge the velocity of stars follows a three-dimensional Gaussian distribution, and in the disc, stars have a constant velocity dispersion from a constant rotation speed. In Han & Gould (1995) and Han & Gould (2003), a bulge and two disc components are considered, but the velocity distribution is similar to that in Zhu17 and is also unrelated to the position. In Bennett et al. (2014), they considered four Galactic components, a bar, a spheroidal stellar halo, a thin and a thick disc (Robin et al. 2003). Even though they considered seven stellar populations, the velocity distribution of each population is still Gaussian and similar to Zhu17 except they set a cutoff escape velocity. In Lam et al. (2020), they made use of Galaxia (Sharma et al. 2011) as their Galactic model. It is more complex and includes not only three disc components but also a spheroidal halo and a bulge. Their stellar dynamics contains many details in the real Milky Way, like radial dependence and asymmetric drift of the disc. For more information, see Table 1-3 in Sharma et al. (2011).

Despite the diverse Galactic model selections, there is little current work that directly compares different models and discusses how they affect the results. Koshimoto & Bennett (2020) and Shan et al. (2019) have done some comparisons, but the models they used are still somehow a combination of the previous simple models. Therefore it is still unknown to what degree simple models are effective and how complex models change or improve the results. In this paper, we introduce two simple but more realistic models, one is a modified Zhu17 model that considers the radial dependence of the velocity dispersions, and the other is based on the Shu distribution function (Shu 1969) which employs the asymmetric drift in the disc velocity distribution. We compare these two models with the standard Zhu17 model to examine the influence of model selection.

The structure of the paper is as follows: The models and some algorithm details are described in the next section. The simulation and Bayesian analysis results are presented in Section 3, and the discussion and conclusion are given in Section 4.

## 2 METHODS

### 2.1 Background

For a microlesing event, the angular Einstein radius is (e.g., Gould 2000b; Mao 2012)

$$\theta_{\rm E} = \sqrt{\kappa M_{\rm L} \pi_{\rm rel}}, \qquad \pi_{\rm rel} = {\rm AU}\left(\frac{1}{D_{\rm L}} - \frac{1}{D_{\rm S}}\right). \qquad (3)$$

The microlensing timescale is the time needed to cross an Einstein radius due to the relative motion between the lens and the source,

$$t_{\rm E} = \frac{\theta_{\rm E}}{\mu_{\rm rel}}, \qquad (4)$$

where $\mu_{\rm rel}$ is the relative proper motion. The microlensing parallax is

$$\boldsymbol{\pi}_{\rm E} = \frac{\pi_{\rm rel}}{\theta_{\rm E}} \hat{\boldsymbol{\mu}}_{\rm rel}, \qquad \pi_{\rm E} = |\boldsymbol{\pi}_{\rm E}| = \frac{\pi_{\rm rel}}{\theta_{\rm E}}, \qquad (5)$$

where $\hat{\boldsymbol{\mu}}_{\rm rel}$ is the direction of the relative proper motion between the lens and the source.

In a specific time interval and for a specific direction, the number of microlensing events (event rate $\Gamma$) is proportional to the angular area $\pi \theta_{\rm E}^2$ and is inversely proportional to the time scale $t_{\rm E}$,

$$\Gamma \propto \frac{\theta_{\rm E}^2}{t_{\rm E}} D_{\rm L}^2 D_{\rm S}^2 \gamma(D_{\rm S}) = \theta_{\rm E} \mu_{\rm rel} D_{\rm L}^2 D_{\rm S}^2 \gamma(D_{\rm S}). \qquad (6)$$

Here $D_{\rm S}^2$ and $D_{\rm L}^2$ are factors that account for the increase of the number of source and lens stars with the distance, and $\gamma(D_{\rm S})$ is a function that describes how the number of detectable sources varies from $D_{\rm S}$ (Kiraga & Paczynski 1994). All simulated events should be weighted by the event rate, for event $i$ the weight is

$$w_i = \Gamma_i \propto \theta_{{\rm E},i} \mu_{{\rm rel},i} D_{{\rm L},i}^2 D_{{\rm S},i}^2 \gamma(D_{{\rm S},i}). \qquad (7)$$

### 2.2 Models

To generate a microlensing event, the information of the source ($D_{\rm S}, \boldsymbol{\mu}_{\rm S}$) and the lens ($D_{\rm L}, \boldsymbol{\mu}_{\rm L}, M_{\rm L}$) is needed, where $\boldsymbol{\mu}_{\rm L}$ and $\boldsymbol{\mu}_{\rm S}$ are the proper motion of the lens and source, respectively. This information can be retrieved through a given Galactic model. A Galactic model can be described in three parts, the density profile, the velocity distribution, and the mass function (MF). The Galactic density model we adopt basically follows Zhu et al. (2017). The density profile includes two components, a bar (bulge) and a disc. The bar's stellar number density follows (Kent et al. 1991; Dwek et al. 1995)

$$n_{\rm B} = n_{\rm B,0}\, e^{-\frac{1}{2}\sqrt{\left[\left(\frac{x'}{x_0}\right)^2 + \left(\frac{y'}{y_0}\right)^2\right]^2 + \left(\frac{z'}{z_0}\right)^4}}, \qquad (8)$$

where $x'$, $y'$ and $z'$ are the positions along the three major axes of the bar, and $x_0, y_0, z_0$ are the scale length along each axis. The bar angle is taken to be $30°$ (Cao et al. 2013; Wegg & Gerhard 2013). All the parameters and their values are shown in Table 1. The disc's stellar density is exponential,

$$n_{\rm D} = n_{\rm D,0} e^{-\frac{R-R_0}{R_{\rm d}} - \frac{|z|}{z_{\rm d}}}. \qquad (9)$$

Here $R$ and $z$ are the galactic cylindrical coordinate components, $R_{\rm d}$ and $z_{\rm d}$ are the scale length along each axis. $R_0 = 8.3$ kpc is the position of solar system (Gillessen et al. 2009). The density profile is axisymmetric since it does not depend on the azimuthal angle $\phi$. The coefficient $n_{\rm D,0}$ is determined by the local stellar density $0.14$ stars/pc$^3$.

The velocity distributions are different for the two components. In the bulge, the velocity follows a three-dimensional Gaussian distribution,

$$V_{\rm B} \sim \exp\left(-\frac{V_x^2}{2\sigma_{{\rm B},x}^2} - \frac{V_y^2}{2\sigma_{{\rm B},y}^2} - \frac{V_z^2}{2\sigma_{{\rm B},z}^2}\right), \qquad (10)$$

where the velocity dispersion $(\sigma_{{\rm B},x}, \sigma_{{\rm B},y}, \sigma_{{\rm B},z}) = (120, 120, 120)$ km/s (Zhu et al. 2017).

For the velocity distribution of the disc, we introduce three models listed below.

1) The standard model: This model is the same as the original Zhu et al. (2017) model except for some updates in parameter values. The

velocity distribution of disc stars at any distance follows a Gaussian distribution,

$$V_\phi \sim N(\overline{V}_\phi, \sigma^2_{D,\phi}), \quad V_z \sim N(\overline{V}_z, \sigma^2_{D,z}),$$

where the velocity dispersions $\sigma_{D,\phi}$ and $\sigma_{D,z}$ are constant at any location, and the values are taken to be solar neighborhood values, $(\overline{V}_\phi, \overline{V}_z) = (220, 0)$ km/s and $(\sigma_{D,\phi}, \sigma_{D,z}) = (33, 18)$ km/s.

2) Model B: The mean velocity is still $\overline{V}_\phi = 220$ km/s and $\overline{V}_z = 0$ km/s anywhere, but the velocity dispersion is a function of the distance from Galactic center $R$,

$$\sigma_{D,(\phi,z)}(R) = \sigma_{D,(R,\phi,z)}(R_0) e^{-\frac{R-R_0}{4R_d}},$$

where $R_0 = 8.3$ kpc is the position of Sun, and the solar neighborhood values are $\sigma_{D,R}(R_0) = 38$ km/s, $\sigma_{D,\phi}(R_0) = 33$ km/s and $\sigma_{D,z}(R_0) = 18$ km/s. The velocity dispersion scale length, $4R_d \sim 10$ kpc, is inferred from a sample of giant stars (Gaia Collaboration et al. 2018b). However, most microlenses are dwarf stars, which may have a different scale length. Nevertheless, we adopt the *Gaia* scaling here; we will return to this point briefly in the discussion.

3) Model C: We use the dynamic model of Shu (1969). The disc model introduces both radial dependence and asymmetric drift. We use software `galpy` [1] (Bovy 2015) to generate the $V_\phi$ distribution function. The $V_\phi$ distribution as a function of $R$ is shown in Fig. 1. The velocity distribution along the $z$ axis is the same as that in Model B. Note that we only use the $V_\phi$ velocity distribution of Shu's model.

The standard model is widely used in the microlensing field since its good simplicity, however, it is unrealistic and dynamically self-inconsistent. Model B and C introduced radial dependence and asymmetric drift to the Galactic model. Even though the other two models are still simple, they are sufficient for us to study the impact of different model selections.

For the mass function, here we first fix $M_L = 0.5 M_\odot$ for all models. From Eqs. 3 and 4 we know that a specific $M_L$ value is just a coefficient of $t_E$ or a translation term of $\log t_E$ ($M_L \to M'_L$, $t'_E \to \sqrt{\frac{M'_L}{M_L}} t_E$, or $\log t'_E \to \log t_E + \frac{1}{2}\log\frac{M'_L}{M_L}$). A continuous $M_L$ distribution will broaden the $t_E$ distribution, but the effect is the same for all models, and will not change the asymptotic behavior at $t_E \ll 10$ d and $t_E \gg 200$ d. The nature of the velocity distribution difference would be buried if we use a continuous MF, therefore in order to control the variables and focus on the influence of model difference, we choose a $\delta$ MF in the first part (Section 3.1). After that, when doing the Bayesian analysis (Section 3.3), $M_L$ is also a parameter that needs to be obtained, and the mass function must be specified as a prior. We will describe our choice of the mass function at that time.

### 2.3 Algorithm

For specific values, we assume that the observation is toward the center of Baade's Window (R.A., Dec.)$_{J2000}$ = ($18^h03^m32.14^s, -30°02'6.96''$) or $(l, b) = (1.02°, -3.92°)$, and we adopt $\gamma(D_S) = D_S^{-2}$ here (Kiraga & Paczynski 1994). Baade's Window is near the Galactic center, so the line of sight is almost on the disc plane, therefore in Galactic coordinates, we take

$$V_l \simeq V_\phi \cos(l+\theta) + V_R \sin(l+\theta), \quad V_b \simeq V_z, \quad (11)$$

[1] http://github.com/jobovy/galpy

where $l$ is the Galactic longitude and $\theta$ is the angle between the Galactic center to the Sun and to the star. We can easily covert the generated velocity to observed relative angular velocities by

$$\boldsymbol{\mu}_L = \frac{\boldsymbol{V}_L - \boldsymbol{V}_\odot}{D_L}, \quad \boldsymbol{\mu}_S = \frac{\boldsymbol{V}_S - \boldsymbol{V}_\odot}{D_S}, \quad (12)$$

and

$$\boldsymbol{\mu}_{\text{rel}} = \boldsymbol{\mu}_L - \boldsymbol{\mu}_S, \quad \mu_{\text{rel}} = |\boldsymbol{\mu}_{\text{rel}}|. \quad (13)$$

Here $\boldsymbol{V}_L = V_{L,l}\hat{e}_l + V_{L,b}\hat{e}_b$ and $\boldsymbol{V}_S = V_{S,l}\hat{e}_l + V_{S,b}\hat{e}_b$ are the lens and source velocities, respectively, and $\boldsymbol{V}_\odot$ is the velocity of the sun.

In practice, we first randomly generate a pair of lens and source position $D_L$ and $D_S$ from the stellar density profile. To make the computation more efficient, we adjust the stellar density profile to

$$n'_S \propto (n_B + n_D) D_S^2 \gamma(D_S), \quad n'_L \propto (n_B + n_D) D_L^2, \quad (14)$$

where $n'_S$ and $n'_L$ represent the adjusted source and the lens "number density". Using this new density profile to generate $D_L$ and $D_S$ can avoid a large number of very low weight (e.g., small $D_L$) events and improve the computing efficiency.

Second, we generate their velocities according to a given Galactic model's velocity distribution. For the standard model, the generation of velocity is straightforward. For Model B, we calculate the velocity dispersion $\sigma_\phi(R)$ and $\sigma_z(R)$ every 0.1 kpc, and use the nearest value of each lens or source for velocity generation. For Model C, the random number generator in `galpy` is time consuming, so we calculated a discrete sampled $v_\phi$ distribution PDF at specific $R$ every 0.1 kpc, and then linearly interpolated the results. The sample interval is $V_c/200$, or 1.1 km/s, it is precise enough to show the difference between these models.

Then, we calculate the timescale $t_E$ and the weight through Eqs. 3 and 4. Note that we have already taken $D_L^2 D_S^2 \gamma(D_S)$ term when generating their distance, so the weight of an event $i$ is now simply

$$w_i = \theta_{E,i} \mu_{\text{rel},i}. \quad (15)$$

Finally, we repeat the process (which can easily be parallelized) to simulate a large number of events to obtain the statistical properties that we are interested in.

## 3 RESULTS

### 3.1 Timescale Distribution

We generated $\sim 4 \times 10^{10}$ events for each model. Then the results are binned and normalized in $\log t_E$ space by their weight to obtain the $t_E$ distribution.

The first row of Fig. 2 shows the normalized $t_E$ distribution of each model, the different colours represent different components. Obviously, each Galactic component contributes differently to the distribution. For the standard model, the disc-source disc-lens (D-D) distribution shows a significant double bump feature, and the reason is that the rotation direction of the disc changes at the two sides of the Galactic center. For Models B and C, the rotation direction changes as well, but the second bump is smoothed by their large velocity dispersions near the Galactic center.

The three models also share some common features. Short and long $t_E$ are dominated by bulge-source bulge-lens (B-B) and bulge-source disc-lens (B-D) events, respectively. Short events are dominated by those where lens and source are close, whereas long events are dominated by those have slow relative proper motions. The former mostly exists in the high-density bulge region, while the latter mostly exists in the dynamically colder Galactic disc. Because of the higher stellar density and relatively faster motion (leading to shorter $t_{\rm E}$'s) in the bulge region, bulge-source events contribute more than 80% of all microlensing events in all of the three models, which is consistent with the estimation of Kiraga & Paczynski (1994).

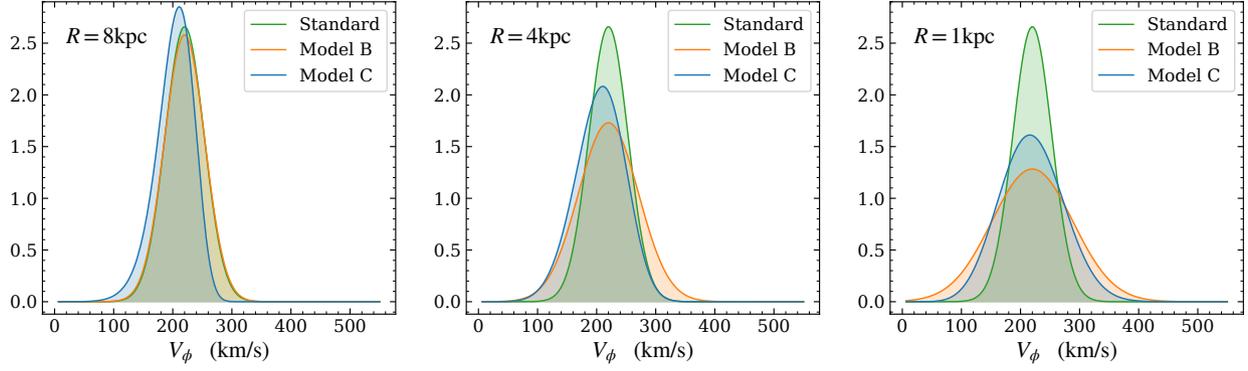

**Figure 1.** The $V_\phi$ distribution of three models at a given distance ($R$) from the Galactic center. Notice that the local rotation speed is fixed at 220 km/s and the Sun is at $R$ = 8.3 kpc.

**Table 1.** Parameters adopted in the simulation

| Parameter | Value | Unit | Remarks | Reference |
|---|---|---|---|---|
| $R_0$ | 8.3 | kpc | Distance to Galactic center | Gillessen et al. (2009) |
| $\alpha_{\rm bar}$ | 30 | deg | Bar angle | Cao et al. (2013); Wegg & Gerhard (2013) |
| $(x_0, y_0, z_0)$ | (1.59, 0.424, 0.424) | kpc | Bar scale length along each axis | Robin et al. (2003) |
| $n_{\rm B,0}$ | 13.7 | stars/pc$^3$ | Bulge stellar density coefficient | Robin et al. (2003) |
| $(\sigma_{\rm B,x}, \sigma_{\rm B,y}, \sigma_{\rm B,z})$ | (120, 120, 120) | km/s | Bulge velocity dispersion | Zhu et al. (2017) |
| $(R_{\rm d}, z_{\rm d})$ | (2.5, 0.325) | kpc | Disc scale length and scale height | Binney & Tremaine (2008) |
| $V_c$ or $\overline{V}_\phi(R_0)$ | 220 | km/s | Disc mean rotation velocity | Binney & Tremaine (2008) |
| $\overline{V}_z(R_0)$ | 0 | km/s | Disc mean perpendicular velocity | |
| $\sigma_{\rm D,R}(R_0)$ | 38 | km/s | Solar neighbor velocity dispersion | |
| $\sigma_{\rm D,\phi}(R_0)$ | 33 | km/s | Solar neighbor velocity dispersion | |
| $\sigma_{\rm D,z}(R_0)$ | 18 | km/s | Solar neighbor velocity dispersion | |
| $(V_{\odot R}, V_{\odot \phi}, V_{\odot z})$ | $(-11, V_c + 12, 7)$ | km/s | Motion of solar system | Schönrich et al. (2010) |
| $n_{\rm D,0}$ | 0.14 | stars/pc$^3$ | Local stellar number density | |
| $D_{\rm S,B}$ | [5.6, 11.0] | kpc | Distance range of bulge component | |
| $D_{\rm S,D}$, $D_{\rm L,D}$ | [0.01, 16.6] | kpc | Distance range of disc component | |
| $M_{\rm L}$ | 0.5 | $M_\odot$ | Lens mass | |
| $\gamma$ | $D_{\rm S}^{-2}$ | - | The factor of detectable source star numbers when varying the distance | |

The comparisons of different models are shown in Fig. 3. Notice that the the total event rate is not the same, so in this figure only the distribution of standard model is normalized to unity. The total event rate ratio is ∼ 1 : 1.015 : 0.999. The detailed event rate comparison is shown in the figure and Table 2. Because all the $t_{\rm E}$ distributions follow $P(< t_{\rm E}) \propto t_{\rm E}^3$ at the small $t_{\rm E}$ tail and $P(> t_{\rm E}) \propto t_{\rm E}^{-3}$ at the long $t_{\rm E}$ tail (Mao & Paczynski 1996), the ratios of the three models on the log scale tend to be constant at both small and large $t_{\rm E}$. This can be seen as follows: if the $t_{\rm E}$ distribution for model $m$ is a power-law, for example, the $t_{\rm E} \to \infty$ part that $p_m(\log t_{\rm E}) = A_m t_{\rm E}^{-3}$, the fraction of $> t_{\rm E}$ events will be

$$P_m(> t_{\rm E}) = \int_{t_{\rm E}}^{\infty} p_m(\log t_{\rm E}) \, {\rm d}\log t_{\rm E} = \frac{\ln 10}{3} A_m t_{\rm E}^{-3}. \quad (16)$$

So $P_1(> t_{\rm E}) : P_2(> t_{\rm E}) = p_1(\log t_{\rm E}) : p_2(\log t_{\rm E}) = A_1 : A_2$, we can read the event rate ratio between two models directly from the log-plot. The relation is also on small $t_{\rm E}$ power-law tails.

At the $t_{\rm E} > 300$ days tail, Models B and C only produce ∼ 84% and ∼ 95% events of the standard model. In other words, if an individual event shows long $t_{\rm E}$, the lens would be slightly more likely to be a massive object, because in these two models, low-mass objects cannot produce as many long $t_{\rm E}$ events as in the standard model. In addition, at the $t_{\rm E} < 3$ days tail, Model B and Model C produce a little more (∼ 5.0% and 3.5%) events than the standard model. However, for the overall $t_{\rm E}$ distributions, the difference may be not significant for the current microlensing database since these extreme events only accounted for less than 1%.

### 3.2 $\theta_{\rm E}$ and $\pi_{\rm E}$ Distribution

From Eqs. 3 and 5, we can see that $\theta_{\rm E}$ and $\pi_{\rm E}$ are both a function of $M_{\rm L}$, $D_{\rm L}$ and $D_{\rm S}$ but unrelated to the velocity distribution. However, the event rate (Eq. 6) depends on the velocity distribution, thus under our three models, the event rate distribution as a function of $\theta_{\rm E}$ and $\pi_{\rm E}$ would also be different.

In Fig. 2, the second and third rows show the $\theta_{\rm E}$ and $\pi_{\rm E}$ distribution, respectively. The shapes of $\theta_{\rm E}$ and $\pi_{\rm E}$ distributions are similar (actually they are strictly the same) simply because $\theta_{\rm E} = \kappa M_{\rm L} \pi_{\rm E}$ (Eq. 1) and we fix the $M_{\rm L}$ in the simulation. In all models, small $\theta_{\rm E}$ and small $\pi_{\rm E}$ are dominated by bulge-source bulge-lens (B-B) events since these events have small $\pi_{\rm rel}$, and large $\theta_{\rm E}$ and large $\pi_{\rm E}$ are dominated by bulge-source disc-lens (B-D) events because of their large $\pi_{\rm rel}$. The comparisons of the three models are shown in Fig. 4.

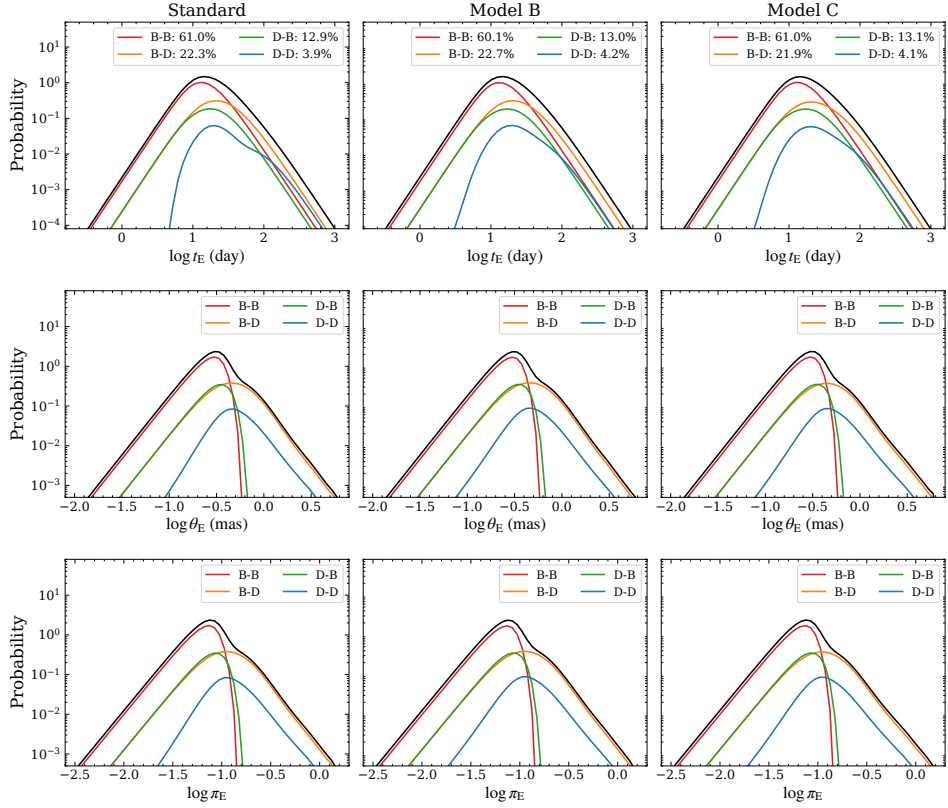

**Figure 2.** The contribution of each component to the $t_E$, $\theta_E$ and $\pi_E$ distributions for each model. Note that we fix $M_L = 0.5 M_\odot$. Different components are shown with different colours. For example, "B-D" represents the bulge-source and disc-lens events, "D-B" represents the disc-source and bulge-lens events, and so on. The black lines represent the overall distribution. The labels in subsequent figures are the same.

**Table 2.** Event Rate Comparison

| Model    | B-B   | B-D   | D-B   | D-D    | All   |
|----------|-------|-------|-------|--------|-------|
| Standard | -     | -     | -     | -      | -     |
| Model B  | +0.0% | +3.3% | +2.5% | +10.8% | +1.5% |
| Model C  | +0.0% | −1.9% | +1.4% | +4.9%  | −0.1% |

**Table 3.** Asymptotic Behaviors of the Total Event Rate

| Model    | $t_E \to 0$ | $t_E \to \infty$ | $\theta_E \to 0$ | $\pi_E \to 0$ |
|----------|-------------|------------------|------------------|---------------|
| Standard | -           | -                | -                | -             |
| Model B  | +4.7%       | −16%             | +0.9%            | +0.9%         |
| Model C  | +3.3%       | −5%              | +0.8%            | +0.8%         |

B-B events are the same in these three models, so the distribution of $\theta_E < 0.2$ mas and $\pi_E < 0.06$ are mostly the same, except that Models B and C produce 1% more events here since the discs are dynamically hotter. On the other hand, at $\theta_E > 3$ mas and $\pi_E > 0.8$ part, the two new models predict more events than the standard model. These parts of the distributions are dominated by small $D_L$ events (leading to large $\pi_{rel}$ according to Eq. 3). For these nearby events, Models B and C have warmer dynamics, so the event rate is a little higher than that in the standard model.

However, these tail events are rare and only contribute $< 0.1\%$ of all events, and they do not have much influence on the statistical results. For specific individual events have extremely small or large $\theta_E$ or $\pi_E$, the model selection might bring an up to $\sim 10\%$ difference in the parameter space.

In addition, from Fig. 4 we find that, similar to $t_E$, $\theta_E$ and $\pi_E$ also follow the power-law asymptotic behavior in the tails, $p(\log \theta_E) \propto \theta_E^3$ and $p(\log \pi_E) \propto \pi_E^3$ when they are very small, and $p(\log \theta_E) \propto \theta_E^{-3}$ and $p(\log \pi_E) \propto \pi_E^{-3}$ when they take large values. The identical asymptotic power-law indices of $(\theta_E, \pi_E)$ and $t_E$ are somewhat a coincidence because $t_E$ depends on kinematics while $(\theta_E, \pi_E)$ do not, and they are dominated by different populations. With a simple toy model as in Mao & Paczynski (1996), we can analytically derive the $\pm 3$ powers (for more details, see the appendix).

### 3.3 Bayesian Analysis

Below we use three examples to illustrate the impact of Galactic models on individual events, where 1) only $t_E$ is available, 2) $t_E$ and $\pi_E$ are available, 3) $t_E$, $\theta_E$ and source proper motion are available.

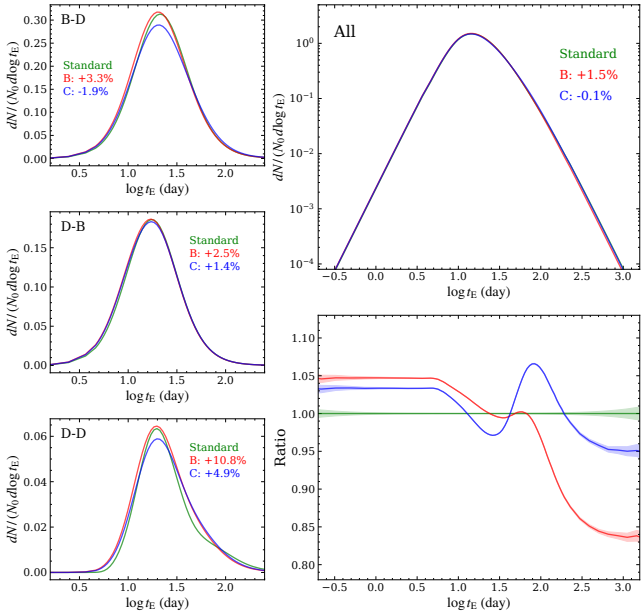

**Figure 3.** The event rate as a function of $t_E$ of the three models and their ratios (bottom right). $N_0$ is the total event rate of the standard model. Here we fix $M_L = 0.5 M_\odot$. Different Galactic components are shown in different panels. Each model is presented in different colours, and the event rate ratio of each component compared to that in the standard model is also shown in the bottom right panel. The ratio of the standard model (green) is normalized to unity, and the total probability of the standard curve (upper right panel) is unity. The colored regions on the lower right panel represent the $1\sigma$ ranges which are from the statistical uncertainties due to the finite sample size ($\sim 4 \times 10^{10}$).

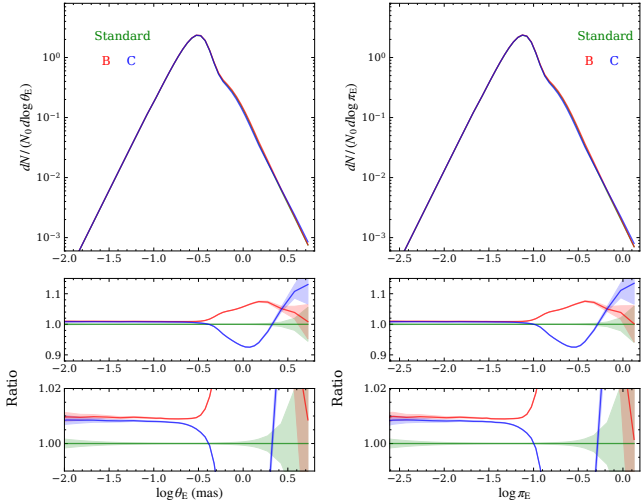

**Figure 4.** The event rate as a function of $\theta_E$ and $\pi_E$ for the three models and their ratios to the standard model. $N_0$ is the total event rate of the standard model. Here we fix $M_L = 0.5 M_\odot$. Each model is presented in a different colour. The ratio of the standard model (green) is normalized to unity, and the total probability of the standard curves are unity. Notice that the ratios at large $\theta_E$ and $\pi_E$ are the same. The colored regions on the ratio plots represent the $1\sigma$ ranges which are from the statistical uncertainties due to the finite sample size ($\sim 4 \times 10^{10}$).

### 3.3.1 $t_E$ is the Only Observable

In reality, for an individual event, the timescale value and its error are obtained by modeling the observation light curve. As mentioned above, if $t_E$ is the only observable, other parameters like $M_L$ and $D_L$ can only be obtained by Bayesian analysis. The Bayesian analysis here is actually an inverse problem of calculating the $t_E$ distribution in Section 3.1, the difference is that the weight of each event $w_i$ needs an extra $t_E$ prior,

$$w_i = \Gamma_i \mathcal{L}_i(t_E), \quad \mathcal{L}_i(t_E) = \frac{\exp\left[-(t_{E,i} - t_E)^2 / 2\sigma_{t_E}^2\right]}{\sqrt{2\pi}\sigma_{t_E}}, \quad (17)$$

where $t_E$ and $\sigma_{t_E}$ is the mean value and standard deviation of the timescale measurement from the observation, and $\mathcal{L}_i(t_E)$ is the likelihood of $t_E$ (assumed to be Gaussian). Here we directly give some values $t_E = 2, 40, 1000$ days and $\sigma_{t_E} = 0.1 t_E$.

Here we use the Kroupa mass function for brown dwarfs and main sequence stars (Kroupa 2001), where

$$\frac{d\xi(M_L)}{d \log M_L} \propto \begin{cases} M_L^{0.7}, & 0.013 < M_L/M_\odot < 0.08 \\ M_L^{-0.3}, & 0.08 < M_L/M_\odot < 0.5 \\ M_L^{-1.3}, & 0.5 < M_L/M_\odot < 1.3 \end{cases}. \quad (18)$$

In addition, we also consider the distribution of white dwarfs (WD), neutron stars (NS) and black holes (BH) (Karolinski & Zhu in prep.), where

$$\frac{d\xi_{WD}}{dM_L} \propto \mathcal{N}(0.65 M_\odot, 0.16^2 M_\odot^2), 0.3 < \frac{M_L}{M_\odot} < 1.4, \quad (19)$$

$$\frac{d\xi_{NS}}{dM_L} \propto \mathcal{N}(1.5 M_\odot, 0.2^2 M_\odot^2), 1.1 < \frac{M_L}{M_\odot} < 2.5, \quad (20)$$

$$\frac{d\xi_{BH}}{dM_L} \propto 10^{-\frac{M_L}{17 M_\odot}}, 2.5 < \frac{M_L}{M_\odot} < 80, \quad (21)$$

and their number densities are taken to be 3.1%, 0.2% and 0.1% of main sequence stars. The fractions and parameters are derived from or extensions of Giammichele et al. (2012) (WD), Kiziltan et al. (2013) and Sartore et al. (2010) (NS), and Olejak et al. (2020) (BH).

We generate $\sim 2 \times 10^{10}$ events for each Galactic model, and then substitute different $t_E$ prior to obtain the $D_L$ and $M_L$ posterior probability distributions (see Figs. 5 and 6). We can see that for the same model, longer $t_E$ decreases the bulge-source probability and increases the disc-source probability. This is consistent with the result in Section 3.1, for which disc-source events contribute more at long $t_E$.

The differences of $D_L$ and $M_L$ between different models is only a few percent and mostly emerges on long $t_E$ events. Table 4 shows the detailed parameters for an event with $t_E = 1000 \pm 100$ days. Only D-D components show apparent differences among different models. But the differences are still small compare to their $1\sigma$ uncertainties. Moreover, such extreme events are rare in reality, which indicates that the Bayesian analysis is insensitive to Galactic models if $t_E$ is the only observable. Thus the conclusions made by previous work with simple models may still be valid.

### 3.3.2 A Black Hole Candidate: OGLE3-ULENS-PAR-02

Here we apply the Bayesian analysis procedure on a black hole candidate OGLE3-ULENS-PAR-02 found by Wyrzykowski et al. (2016). The coordinate of this event is $(RA, Dec)_{J2000} = (17^h 57^m 23.14^s, -28°46'32.0'')$. The light-curve analyses show the parameters are (we choose the $u_0 > 0$ solution here): $t_E = 296.1^{+7.6}_{-7.4}$ days, $\pi_{E,E} = -0.051^{+0.002}_{-0.002}$, and $\pi_{E,N} = 0.033^{+0.001}_{-0.001}$, where $\pi_{E,E}$ and $\pi_{E,N}$ are the eastern and northern components of $\pi_E$.

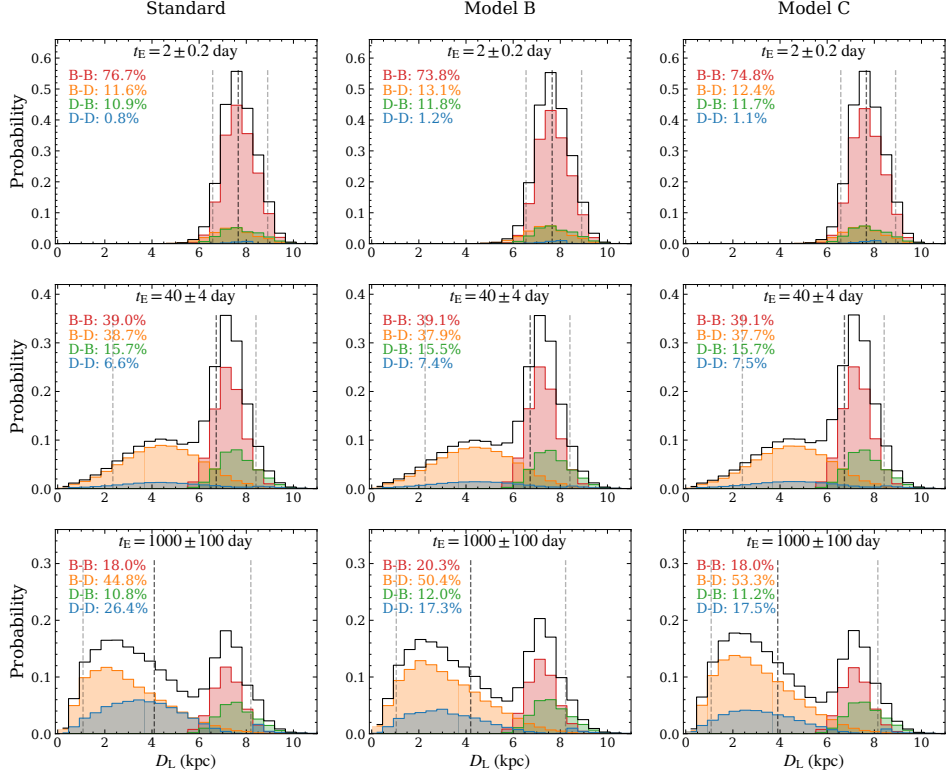

**Figure 5.** The posterior probability distribution of $D_L$ under the given Galactic model and $t_E$ prior. Each column represents a Galactic model, and each row uses the same $t_E$ prior. The vertical dashed lines denote the mean and 90% confidence interval of the overall distribution. The fraction of each component (B-B, B-D, D-B, D-D) is shown on each panel, too.

**Table 4.** Bayesian Results of an event with the only observable $t_E = 1000 \pm 100$ days

| Model\Component | | B-B | B-D | D-B | D-D | All |
|---|---|---|---|---|---|---|
| Standard | $P_{\rm Comp}$ | 18.0% | 44.8% | 10.8% | 26.4% | – |
| | $D_L$ | $7.2^{+0.6}_{-0.6}$ kpc | $2.7^{+2.0}_{-1.3}$ kpc | $7.6^{+0.9}_{-0.7}$ kpc | $3.6^{+1.8}_{-1.7}$ kpc | $4.1^{+4.1}_{-3.0}$ kpc |
| | $M_L$ | $9.2^{+18.5}_{-8.6}$ $M_\odot$ | $7.7^{+18.2}_{-7.7}$ $M_\odot$ | $8.3^{+18.4}_{-7.1}$ $M_\odot$ | $1.2^{+19.5}_{-0.7}$ $M_\odot$ | $5.8^{+31.9}_{-5.5}$ $M_\odot$ |
| | $P(>2.5M_\odot)$ | 57.2% | 54.9% | 55.7% | 43.7% | 52.5% |
| Model B | $P_{\rm Comp}$ | 20.3% | 50.4% | 12.0% | 17.3% | – |
| | $D_L$ | $7.2^{+0.6}_{-0.6}$ kpc | $2.9^{+2.0}_{-1.4}$ kpc | $7.6^{+0.9}_{-0.7}$ kpc | $3.1^{+2.0}_{-1.5}$ kpc | $4.2^{+4.0}_{-3.2}$ kpc |
| | $M_L$ | $9.1^{+18.5}_{-8.5}$ $M_\odot$ | $7.7^{+18.7}_{-7.1}$ $M_\odot$ | $8.5^{+18.3}_{-7.9}$ $M_\odot$ | $4.1^{+20.2}_{-3.5}$ $M_\odot$ | $7.6^{+31.7}_{-7.3}$ $M_\odot$ |
| | $P(>2.5M_\odot)$ | 56.9% | 55.0% | 56.0% | 50.8% | 54.8% |
| Model C | $P_{\rm Comp}$ | 18.0% | 53.3% | 11.2% | 17.5% | – |
| | $D_L$ | $7.2^{+0.6}_{-0.6}$ kpc | $2.8^{+1.9}_{-1.3}$ kpc | $7.6^{+0.9}_{-0.7}$ kpc | $3.1^{+1.9}_{-1.5}$ kpc | $3.9^{+4.2}_{-2.8}$ kpc |
| | $M_L$ | $8.9^{+18.5}_{-8.3}$ $M_\odot$ | $7.1^{+18.7}_{-6.6}$ $M_\odot$ | $8.7^{+18.6}_{-8.1}$ $M_\odot$ | $4.1^{+20.1}_{-3.5}$ $M_\odot$ | $7.3^{+31.1}_{-7.0}$ $M_\odot$ |
| | $P(>2.5M_\odot)$ | 56.8% | 54.1% | 56.4% | 50.8% | 54.3% |

In this case, the weight for each simulated event becomes

$$w_i = \Gamma_i \, \mathcal{L}_i(t_E) \, \mathcal{L}_i(\pi_{E,E}) \, \mathcal{L}_i(\pi_{E,N}), \quad (22)$$

where $\mathcal{L}_i(t_E)$, $\mathcal{L}_i(\pi_{E,E})$ and $\mathcal{L}_i(\pi_{E,N})$ are the likelihood of $t_E$, $\pi_{E,E}$ and $\pi_{E,N}$[2]. We use a log-uniform mass function in $M_L \in [10^{-2}, 10^3] M_\odot$ as the $M_L$ prior in all models. We simulated $\sim 6 \times 10^{10}$ events and the results are shown in Table 5 and Fig. 7. We find the disc-lens probability dominates in all models, and the masses are $6.3^{+5.1}_{-3.8} M_\odot$, $6.8^{+5.3}_{-4.0} M_\odot$ and $6.9^{+4.9}_{-3.8} M_\odot$ for the standard model, Model B and C, respectively.

[2] We ignore the correlation between $\pi_{E,E}$ and $\pi_{E,N}$.

In Table 5, for disc-lens events, both Model B and Model C estimate a larger mass than the standard model. This is consistent with our discussion in Section 3.1, that under a specific lens mass, Model B and Model C cannot produce as many long $t_E$ events as the standard model, therefore the Bayesian analysis for a given long $t_E$ event tends to be more massive. In addition, the largest $M_L$ part is dominated by B-D and D-D events whichever model we choose. And the fraction of the disc-source component increases in both Model B and Model C, which leads to the higher estimation of $M_L$.

Note that even though the model used in Wyrzykowski et al. (2016) is similar to our standard model, the parameters are quite different, e.g., the velocity dispersion along the azimuthal axis and the z-axis are (40, 55) km/s which are larger than those in our standard model. In addition, they assumed the event is 100% a B-D event. Thus the mass they estimated $M_L = 9.3^{+8.7}_{-4.3} M_\odot$ is different from ours (but still within $1\sigma$), which also proves the importance of model selection.

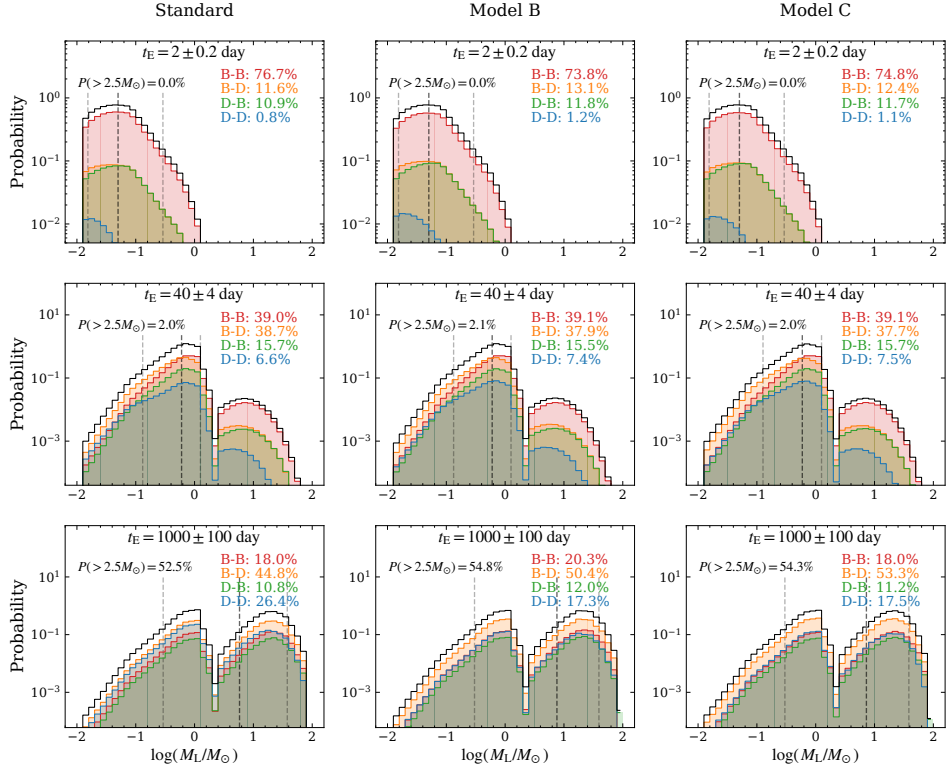

**Figure 6.** The posterior probability distribution of $M_L$ under the given Galactic model and $t_E$ prior. Each column represents a Galactic model, and each row uses the same $t_E$ prior. The vertical dashed lines denote the mean and 90% confidence interval of the overall distribution. The fraction of each component (B-B, B-D, D-B, D-D) and the $> 2.5 M_\odot$ probability are shown on each subfigure, too.

**Table 5.** Bayesian Results of the Black Hole Candidate OGLE3-ULENS-PAR-02

| Model\Component | | B-B | B-D | D-B | D-D | All |
|---|---|---|---|---|---|---|
| Standard | $P_{\text{Comp}}$ | 2.5% | 81.4% | 9.7% | 6.4% | – |
| | $D_L$ | $6.9^{+0.6}_{-0.6}$ kpc | $3.0^{+1.1}_{-0.9}$ kpc | $7.5^{+0.8}_{-0.7}$ kpc | $4.3^{+1.2}_{-1.2}$ kpc | $3.2^{+2.1}_{-1.1}$ kpc |
| | $M_L$ | $1.2^{+0.4}_{-0.4}$ $M_\odot$ | $7.2^{+4.8}_{-3.0}$ $M_\odot$ | $1.8^{+0.5}_{-0.5}$ $M_\odot$ | $2.9^{+2.3}_{-1.3}$ $M_\odot$ | $6.3^{+5.1}_{-3.8}$ $M_\odot$ |
| | $P(>3M_\odot)$ | 0.0% | 93.9% | 2.3% | 47.8% | 79.8% |
| Model B | $P_{\text{Comp}}$ | 2.0% | 79.6% | 7.9% | 10.5% | – |
| | $D_L$ | $6.9^{+0.6}_{-0.6}$ kpc | $2.8^{+1.2}_{-0.9}$ kpc | $7.5^{+0.8}_{-0.7}$ kpc | $3.4^{+1.2}_{-1.1}$ kpc | $3.1^{+1.8}_{-1.0}$ kpc |
| | $M_L$ | $1.2^{+0.4}_{-0.4}$ $M_\odot$ | $7.6^{+5.2}_{-3.4}$ $M_\odot$ | $1.8^{+0.6}_{-0.5}$ $M_\odot$ | $4.7^{+3.7}_{-2.1}$ $M_\odot$ | $6.8^{+5.3}_{-4.0}$ $M_\odot$ |
| | $P(>3M_\odot)$ | 0.0% | 94.0% | 2.4% | 79.2% | 83.3% |
| Model C | $P_{\text{Comp}}$ | 1.6% | 83.4% | 6.9% | 8.1% | – |
| | $D_L$ | $6.9^{+0.6}_{-0.6}$ kpc | $2.8^{+1.1}_{-0.8}$ kpc | $7.5^{+0.8}_{-0.7}$ kpc | $3.6^{+1.1}_{-1.0}$ kpc | $3.0^{+1.6}_{-0.9}$ kpc |
| | $M_L$ | $1.1^{+0.5}_{-0.4}$ $M_\odot$ | $7.7^{+4.7}_{-3.2}$ $M_\odot$ | $1.8^{+0.5}_{-0.5}$ $M_\odot$ | $4.3^{+3.0}_{-1.8}$ $M_\odot$ | $6.9^{+4.9}_{-3.8}$ $M_\odot$ |
| | $P(>3M_\odot)$ | 0.0% | 94.8% | 1.8% | 75.2% | 85.3% |

### 3.3.3 A Free Floating Planet Candidate: OGLE-2017-BLG-0560

We also applied the Bayesian approach to a free-floating planet candidate OGLE-2017-BLG-0560 at the coordinate $(\text{RA}, \text{Dec})_{\text{J}2000} = (17^h51^m51.33^s, -30°27'31.4'')$ (Mróz et al. 2019). This is an ambiguous event because the mass occasionally located at the critical mass between a brown dwarf and a planet ($M_{\text{crit}} \sim 13 M_J$, where $M_J = 0.95 \times 10^{-3} M_\odot$ is the mass of Jupiter). From light-curve modeling and source star identifying, the parameters of this event are yielded: $t_E = 0.905^{+0.005}_{-0.005}$ days, and $\theta_E = 38.7^{+1.6}_{-1.6} \mu\text{as}$. The proper motion of the source star is also available in *Gaia* DR2 (Gaia Collaboration et al. 2018a), where $\mu_{S,\text{RA}} = -2.19 \pm 0.30$ mas/yr, $\mu_{S,\text{Dec}} = -11.7 \pm 0.24$ mas/yr and their correlation $\rho_{\mu_{S,\text{RA}},\mu_{S,\text{Dec}}} = 0.276$.

Here we generate the events from the Galactic model and multiply an extra $\mu_S$ likelihood on the weights. We used a log-uniform mass function for $M_L \in [10^{-5}, 10^1] M_\odot$ and simulated $\sim 10^{11}$ events to acquire the posterior $M_L$ and $D_L$ distribution, the weight of each event in this case is

$$w_i = \Gamma_i \, \mathcal{L}_i(t_E) \, \mathcal{L}_i(\theta_E) \, \mathcal{L}_i(\mu_S), \tag{23}$$

where $\mathcal{L}_i(\theta_E)$ is the likelihood of $\theta_E$ and $\mathcal{L}_i(\mu_S)$ is the likelihood of source proper motion $\mu_S$ which is assumed to be a two-dimensional Gaussian. Table 6 and Fig. 8 show the results in detail.

For the standard model, our result is qualitatively consistent with Mróz et al. (2019). In our case, the lens is either a Jupiter-mass planet in the disc ($M_L = 1.5^{+2.9}_{-0.7}$ $M_J$, $D_L = 4.1^{+1.8}_{-1.2}$ kpc) or a brown dwarf in the bulge ($M_L = 18.1^{+47.0}_{-9.4}$ $M_J$, $D_L = 7.7^{+0.6}_{-0.6}$ kpc), the probability ratio is $P_{\rm bulge} : P_{\rm disc} \approx 79\% : 21\%$. However, Model B and Model C show slightly different results. Bulge-lens events are the same for all of the three models, but Model B and Model C have a dynamically warmer disc, making it easier to produce short $t_E$ events in the disc for a fixed lens mass. So for the bulge-source disc-lens components, the two models show higher lens mass estimation of $M_L = 6.5^{+14.5}_{-4.3} M_J$ and $M_L = 8.9^{+22.3}_{-6.1} M_J$. However, a warmer disc also increases the event rate, leading to a significant higher disc-lens probability ($P_{\rm bulge} : P_{\rm disc} \approx 29\% : 71\%$ in B and 48% : 52% in C) leading to little change on the overall $M_L$ estimation. In Model B, the overall $M_L$ is even a bit smaller. A larger $P_{\rm disc}$ also makes the probability of the lens being a planet $P(<13M_J)$ higher. In addition, for Models B and C, the dynamics of the inner disc is more similar to the bulge population than that in the standard model, thus the estimations of $D_L$ ($6.6^{+1.0}_{-1.8}$ kpc and $7.0^{+0.8}_{-1.7}$ kpc for Models B anc C) show that the lens are more likely to be located in the inner part of the disc.

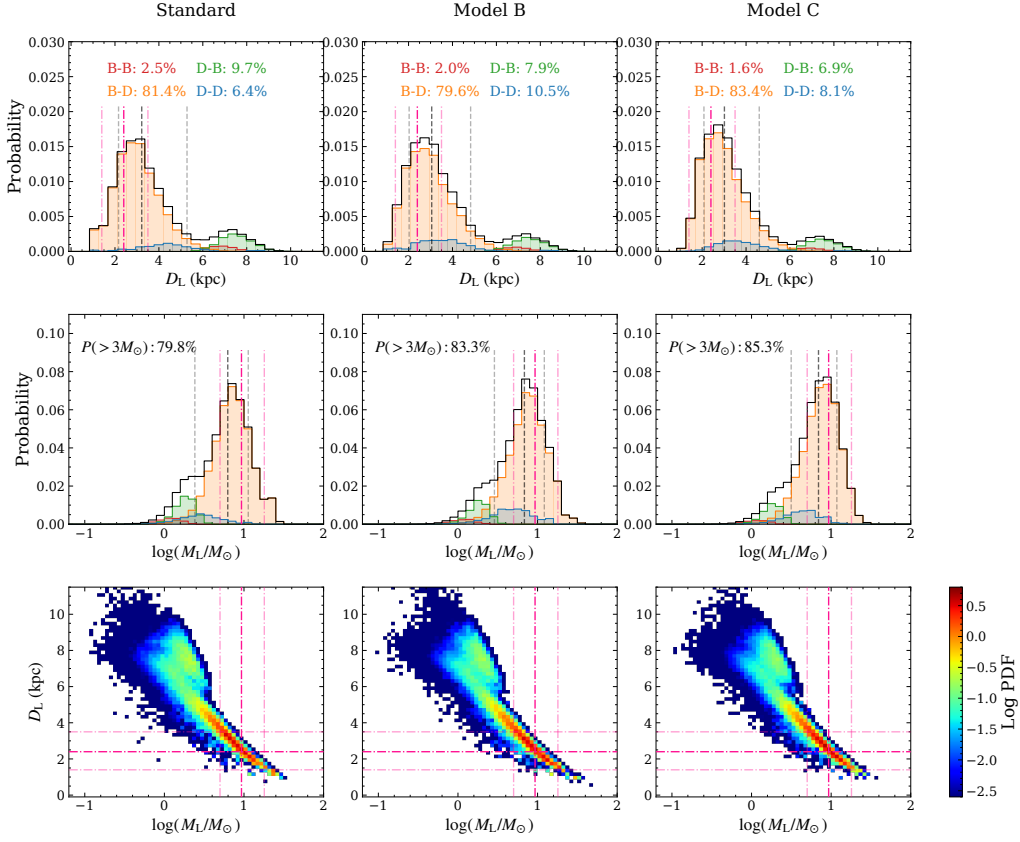

**Figure 7.** The posterior probability distribution of OGLE3-ULENS-PAR-02. The fraction of each Galactic component is labeled on the top of $D_L$ plot, and the probability of $M_L > 3M_\odot$ is shown on the $M_L$ plot. The black vertical dashed lines denote the mean value and 68.3% confidence interval of the overall distribution. We also label the mean and $\pm 1\sigma$ values for $D_L$ and $M_L$ from Wyrzykowski et al. (2016) with the pink dashed lines.

**Table 6.** Bayesian Results of the Free-floating Planet Candidate OGLE-2017-BLG-0560

| Model | \Component | B-B | B-D | D-B | D-D | All |
|---|---|---|---|---|---|---|
| Standard | $P_{\rm Comp}$ | 78.6% | 21.4% | 0.0% | 0.0% | – |
|  | $D_L$ | $7.7^{+0.6}_{-0.6}$ kpc | $4.1^{+1.8}_{-1.2}$ kpc | – | – | $7.5^{+0.7}_{-2.1}$ kpc |
|  | $M_L$ | $18.1^{+47.0}_{-9.4}$ $M_J$ | $1.5^{+2.9}_{-0.7}$ $M_J$ | – | – | $13.8^{+38.4}_{-10.9}$ $M_J$ |
|  | $P(<13M_J)$ | 34.7% | 96.8% | – | – | 48.0% |
| Model B | $P_{\rm Comp}$ | 28.6% | 69.9% | 0.4% | 1.0% | – |
|  | $D_L$ | $7.7^{+0.7}_{-0.6}$ kpc | $6.6^{+1.0}_{-1.8}$ kpc | $8.3^{+0.3}_{-0.7}$ kpc | $7.0^{+0.8}_{-2.6}$ kpc | $7.1^{+0.8}_{-1.8}$ kpc |
|  | $M_L$ | $19.2^{+47.8}_{-10.5}$ $M_J$ | $6.5^{+14.5}_{-4.3}$ $M_J$ | $18.6^{+63.1}_{-9.3}$ $M_J$ | $6.7^{+7.1}_{-4.9}$ $M_J$ | $9.2^{+24.9}_{-6.4}$ $M_J$ |
|  | $P(<13M_J)$ | 32.0% | 73.6% | 29.0% | 82.1% | 61.6% |
| Model C | $P_{\rm Comp}$ | 47.2% | 50.9% | 1.1% | 0.8% | – |
|  | $D_L$ | $7.7^{+0.6}_{-0.6}$ kpc | $7.0^{+0.8}_{-1.7}$ kpc | $8.0^{+0.5}_{-0.5}$ kpc | $6.8^{+1.0}_{-1.5}$ kpc | $7.4^{+0.6}_{-1.2}$ kpc |
|  | $M_L$ | $19.3^{+53.0}_{-10.7}$ $M_J$ | $8.9^{+22.3}_{-6.1}$ $M_J$ | $17.9^{+37.9}_{-7.4}$ $M_J$ | $5.2^{+10.7}_{-3.0}$ $M_J$ | $13.5^{+33.6}_{-8.7}$ $M_J$ |
|  | $P(<13M_J)$ | 31.5% | 63.9% | 26.9% | 82.6% | 48.3% |

## 4 DISCUSSION AND CONCLUSION

The two examples in Sections 3.3.2 and 3.3.3 show the influence of the Galactic model selection on the understanding of individual events. It only slightly changes the estimation of $D_L$ and $M_L$, but often more significantly, the Galactic model affect $P_{\text{bulge-lens}} : P_{\text{disc-lens}}$ (for the importance of the planet bulge-disc ratio, see Zhu et al. 2017). By comparing the Sections in 3.3 we can see, generally, the more observables are measured and more precisely they are, the greater the influence of the Galactic model in Bayesian approach. But the effects on individual events are generally small. However, for statistical works through Bayesian analyses to determine the mass and physical distance (e.g., Cassan et al. 2012), the small differences can accumulate, and the impact may be more significant. For those studies, the Galactic model should be treated with care.

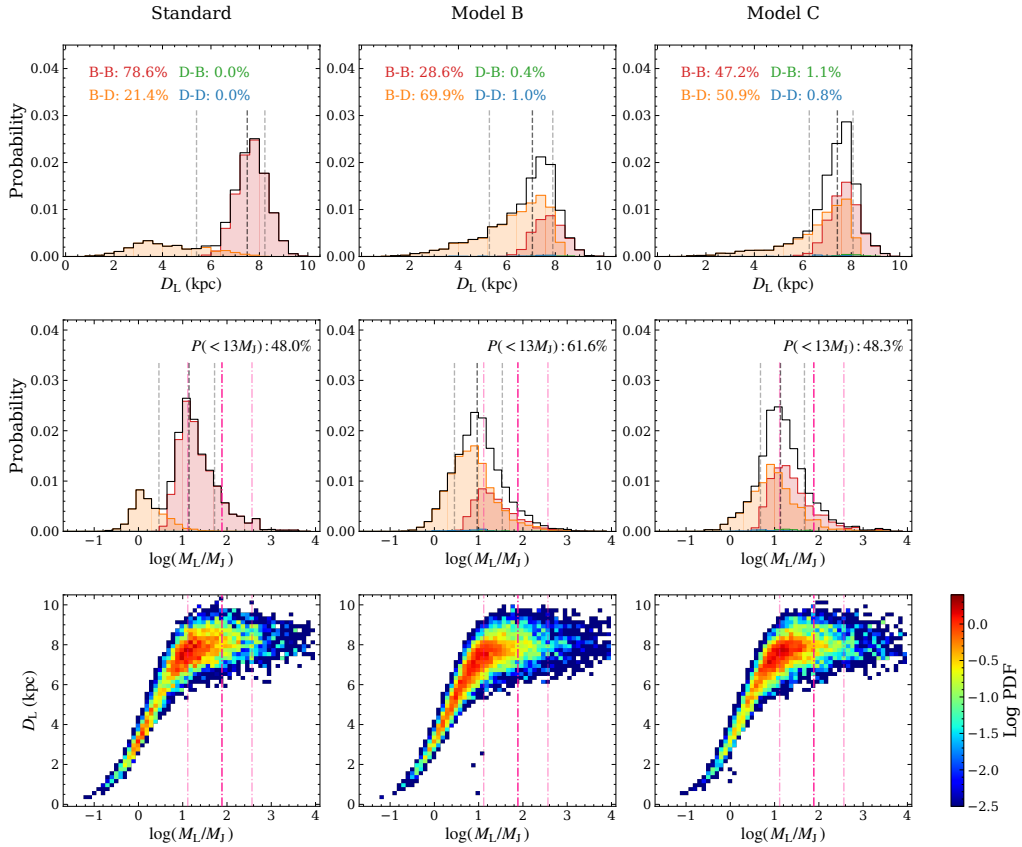

**Figure 8.** The $D_L$, $M_L$ and combined posterior probability distribution of the free-floating planet candidate OGLE-2017-BLG-0560. The fraction of each Galactic component is labeled on the top of $D_L$ plot, and the probability of $M_L < 13 M_J$ is shown on the $M_L$ plot. The black vertical dashed lines denote the mean value and 68.3% confidence interval of the overall distribution. We also label the mean and $\pm 1\sigma$ posterior $M_L$ values from Mróz et al. (2019) (the result of prior 2 in it) with the pink dashed lines.

On the other hand, if enormous enough microlensing events are observed, it is possible to constrain the Galactic model as well. Wood & Mao (2005) tested the Han & Gould (2003) model with the OGLE-II observations (Sumi et al. 2006) by the $t_E$ distribution, and found a consistency at the ~ 52% confidence level. However the sample is small at that time. Shan et al. (2019) used the *Spitzer* microlensing events to test two commonly used Galactic models, and the models showed good consistency. The event sample they used has good constraints on physical parameters, but is of moderate size (includes 13 published *Spitzer* events) and no extreme events (e.g., $t_E > 300$ days or $t_E < 3$ days) are included. The models may show some difference if any "tail" events have happened. The current Galactic model are derived from the visible part of the Milky Way galaxy, while microlensing is also sensitive to dark objects (e.g., stellar remnants, free-floating planets, and perhaps primordial black holes) in the galaxy. They may have different density profiles, velocity distributions and mass functions, and microlensing is the unique way to detect and identify them.

The Galactic model plays a critical role in the Bayesian analysis of microlensing events. Before a feasible way of systematically measuring $\theta_E$ and $\pi_E$ is found, the Bayesian approach is still the main way for us to learn the lens properties. Even though the Bayesian results for individual events are less affected by the Galactic dynamic model, we should be more careful when dealing with a large number of events, because small effects can add up.

In this work, we have considered the two known effects of disk kinematics in our Galaxy, that is the asymmetric drift and the radial variation of the velocity dispersions. We notice that the variation of the velocity dispersion with radius is somewhat uncertain. Under the isothermal sheet assumption, the velocity dispersion should scale as $\exp(-R/2R_d)$, but the *Gaia* data (Gaia Collaboration et al. 2018b) appear to show a more slowly declining trend $\sim \exp(-R/4R_d)$ as inferred from giant stars. This trend has been adopted in this work, but it is unclear whether this applies to microlenses which are mostly dwarf stars, and so we also tried the usual isothermal sheet model. Our qualitative results are not significantly changed. It is reassuring that the differences between the Galactic models do not appear to be significant compared to our current observational uncertainties.

There are other areas that can be improved in the Galactic model. For example, we only considered simple, analytical bar models, while in reality, bars may be more complex (e.g. Wegg et al. 2015). In the future, when more and more events are used to understanding the structure of the Milky Way, more refined models, such as that in `Galaxia` (Sharma et al. 2011), may become necessary.

## ACKNOWLEDGEMENTS

H.Y., S.M., X.Z. and W.Z. acknowledge support by the National Natural Science Foundation of China (Grant No. 11821303 and 11761131004). We thank Wei Zhu for insightful discussions and the referee for a helpful report which improved the paper. We thank the efforts of the `galpy` software developers.

## DATA AVAILABILITY

The data generated in this work will be shared on reasonable requests to the corresponding author.

## APPENDIX A: THE POWER LAW TAIL BEHAVIORS IN $\theta_E$ AND $\pi_E$ DISTRIBUTIONS

Consider the simplest model that the source is at a fixed distance $D_S$, and the lenses are uniformly distributed along the line of sight with constant mass $M$. The lens velocity $V$ is independent of $D_L$. The Einstein radius of a microlensing event is

$$R_E^2 = \frac{4GM}{c^2} \frac{D_L}{D_S}(D_S - D_L) = \frac{4GMD_S}{c^2} x(1-x), \tag{A1}$$

where we define $x \equiv D_L/D_S$. The event rate is

$$d\Gamma \propto \frac{R_E^2}{t_E} n_L(D_L) \, dD_L \, p(V) \, d^3V \propto R_E \, dD_L \, V_\perp p(V) \, d^3V \propto R_E \, dD_L, \tag{A2}$$

where $n_L(D_L)$ is the lens distance distribution which is a constant, $p(V)$ is the velocity distribution, and $V_\perp$ is the transverse velocity. We omit the velocity term because both $\theta_E$ and $\pi_E$ are pure geometric parameters that are independent of $V$. Furthermore, in our simple model, $V$ is also independent of the distance $D_L$, the integration will give a constant. Then the total event rate is given by

$$\Gamma_{\rm tot} = \int_{0<D_L<D_S} d\Gamma = \int_{0<x<1} d\Gamma \propto \int_0^1 \sqrt{x(1-x)} \, dx. \tag{A3}$$

For the $\theta_E$ distribution, we are interested in the probability that $\theta_E < \theta_0$,

$$P(\theta_E < \theta_0) = \frac{1}{\Gamma_{\rm tot}} \int_{\theta_E < \theta_0} d\Gamma. \tag{A4}$$

The $\theta_E < \theta_0$ condition puts constraint on $x$ as

$$P(\theta_E < \theta_0) = P(\sqrt{\kappa M \pi_{\rm rel}} < \theta_0)$$
$$= P(\frac{1}{\beta}\sqrt{\frac{1}{x}-1} < \theta_0) = P\left[x > (\beta^2\theta_0^2+1)^{-1}\right], \tag{A5}$$

where $\pi_{\rm rel} \equiv \frac{1}{D_L} - \frac{1}{D_S} = \frac{1}{D_S}(\frac{1}{x}-1)$ and we define

$$\beta \equiv \sqrt{\frac{D_S}{\kappa M}}. \tag{A6}$$

Therefore

$$P(\theta_E < \theta_0) \propto \int_{(\beta^2\theta_0^2+1)^{-1}}^1 \sqrt{x(1-x)} \, dx$$
$$= \frac{1}{4}\left[\frac{\beta\theta_0(\beta^2\theta_0^2-1)}{(\beta^2\theta_0^2+1)^2} + \arctan(\beta\theta_0)\right]. \tag{A7}$$

Normalizing the result, we find

$$P(\theta_E < \theta_0) = \frac{2}{\pi}\left[\frac{\beta\theta_0(\beta^2\theta_0^2-1)}{(\beta^2\theta_0^2+1)^2} + \arctan(\beta\theta_0)\right]. \tag{A8}$$

Expanding the result at $\theta_0 = 0$, we have

$$P(\theta_E < \theta_0) = \frac{16}{3\pi}(\beta\theta_0)^3 - \frac{48}{5\pi}(\beta\theta_0)^5 + \ldots. \quad (A9)$$

Similarly, for $\theta_0 \to \infty$,

$$P(\theta_E > \theta_0) = 1 - P(\theta_E < \theta_0) = \frac{16}{3\pi}(\beta\theta_0)^{-3} - \frac{48}{5\pi}(\beta\theta_0)^{-5} + \ldots. \quad (A10)$$

Thus we obtain the asymptotic behaviors for the $\log \theta_E$ distribution as follows:

$$p(\log \theta_E)|_{\theta_E \to 0} = \left.\frac{dP(<\theta_E)}{d\log\theta_E}\right|_{\theta_E\to 0} \approx \frac{16\ln 10}{\pi}\beta^3 \theta_E^3; \quad (A11)$$

$$p(\log \theta_E)|_{\theta_E \to \infty} = \left.\frac{dP(<\theta_E)}{d\log\theta_E}\right|_{\theta_E\to \infty} \approx \frac{16\ln 10}{\pi}\beta^{-3} \theta_E^{-3}. \quad (A12)$$

Similarly, for $\pi_E$ we have

$$P(\pi_E < \pi_0) = P(\sqrt{\frac{\pi_{\rm rel}}{\kappa M}} < \pi_0)$$
$$= P(\frac{\beta}{D_S}\sqrt{\frac{1}{x}-1} < \pi_0) = P\left[x > (\frac{D_S^2}{\beta^2}\pi_0^2+1)^{-1}\right], \quad (A13)$$

and

$$P(<\pi_E) \approx \frac{16}{3\pi}\left(\frac{D_S}{\beta}\right)^3 \pi_E^3, \qquad (\pi_E \to 0); \quad (A14)$$

$$P(>\pi_E) \approx \frac{16}{3\pi}\left(\frac{D_S}{\beta}\right)^{-3} \pi_E^{-3}, \qquad (\pi_E \to \infty). \quad (A15)$$

And the asymptotic behaviors for $\log \pi_E$ are given by

$$p(\log \pi_E)|_{\pi_E \to 0} = \left.\frac{dP(<\pi_E)}{d\log\pi_E}\right|_{\pi_E\to 0} \approx \frac{16\ln 10}{\pi}\left(\frac{D_S}{\beta}\right)^3 \pi_E^3; \quad (A16)$$

$$p(\log \pi_E)|_{\pi_E \to \infty} = \left.\frac{dP(<\pi_E)}{d\log\pi_E}\right|_{\pi_E\to \infty} \approx \frac{16\ln 10}{\pi}\left(\frac{D_S}{\beta}\right)^{-3} \pi_E^{-3}. \quad (A17)$$

Although we derived the asymptotic behavior from a simple toy model, the ±3 power-law behaviors on the tails appear to be valid on many other situations. For example, for $P(<\theta_E)$, the coefficient $\beta$ will be a function of the mass $M$ if the mass has a distribution $f(M)$, and the asymptotic behavior at $\theta_E \to 0$ will be

$$P(<\theta_E) \approx \sum_i \frac{16}{3\pi}\beta^3(M_i)f(M_i)\theta_E^3 \propto \theta_E^3, \quad (A18)$$

where the tail behavior $P(<\theta_E) \propto \theta_E^3$ still holds. In Fig. 2, the bulge-lens probability distributions (red and green) do not follow the asymptotic $-3$ power-law behaviors at large $\theta_E$ ($\pi_E$) because the bulge does not extend to the solar neighborhood.

This paper has been typeset from a T<sub>E</sub>X/L<sup>A</sup>T<sub>E</sub>X file prepared by the author.